\documentclass{aa}

\def\msol{\hbox{\kern 0.20em $M_\odot$}}
\def\lsol{\hbox{\kern 0.20em $L_\odot$}}
\def\rsol{\hbox{\kern 0.20em $R_\odot$}}
\def\sr{\hbox{\kern 0.20em sr}}
\def\srmu{\hbox{\kern 0.20em sr$^{-1}$}}
\def\g{\hbox{\kern 0.20em g}}
\def\gmu{\hbox{\kern 0.20em g$^{-1}$}}
\def\kg{\hbox{\kern 0.20em kg}}
\def\pc{\hbox{\kern 0.20em pc}}
\def\mum{\hbox{\kern 0.20em $\mu$m}}
\def\mumd{\hbox{\kern 0.20em $\mu$m$^{-2}$}}
\def\cm{\hbox{\kern 0.20em cm}}
\def\m{\hbox{\kern 0.20em m}}
\def\km{\hbox{\kern 0.20em km}}
\def\nm{\hbox{\kern 0.20em nm}}
\def\s{\hbox{\kern 0.20em s}}
\def\h{\hbox{\kern 0.20em h}}
\def\sec{\hbox{\kern 0.20em sec}}
\def\min{\hbox {\kern 0.20em min}}
\def\smu{\hbox{\kern 0.20em s$^{-1}$}}
\def\smd{\hbox{\kern 0.20em s$^{-2}$}}
\def\an{\hbox{\kern 0.20em an}}
\def\anmu{\hbox{\kern 0.20em an$^{-1}$}}
\def\deg{\hbox{\kern 0.20em $^{\rm o}$}}
\def\yr{\hbox{\kern 0.20em yr}}
\def\yrmu{\hbox{\kern 0.20em yr$^{-1}$}}
\def\Myr{\hbox{\kern 0.20em Myr}}
\def\Mymu{\hbox{\kern 0.20em Myr$^{-1}$}}
\def\K{\hbox{\kern 0.20em K}}
\def\pcmu{\hbox{\kern 0.20em pc$^{-1}$}}
\def\pcmd{\hbox{\kern 0.20em pc$^{-2}$}}
\def\pcmt{\hbox{\kern 0.20em pc$^{-3}$}}
\def\kms{\hbox{\kern 0.20em km\kern 0.20em s$^{-1}$}}
\def\kmpd{\hbox{\kern 0.20em km$^{2}$}}
\def\kpc{\hbox{\kern 0.20em kpc}}
\def\cms{\hbox{\kern 0.20em cm\kern 0.20em s$^{-1}$}}
\def\erg{\hbox{\kern 0.20em erg}}
\def\ergs{\hbox{\kern 0.20em erg}}
\def\cmpd{\hbox{\kern 0.20em cm$^2$}}
\def\cmmd{\hbox{\kern 0.20em cm$^{-2}$}}
\def\cmms{\hbox{\kern 0.20em cm$^{-6}$}}
\def\cmpt{\hbox{\kern 0.20em cm$^3$}}
\def\cmmt{\hbox{\kern 0.20em cm$^{-3}$}}
\def\mpd{\hbox{\kern 0.20em m$^2$}}
\def\mmd{\hbox{\kern 0.20em m$^{-2}$}}
\def\mpt{\hbox{\kern 0.20em m$^3$}}
\def\mmt{\hbox{\kern 0.20em m$^{-3}$}}
\def\mujy{\hbox{\kern 0.20em $\mu$Jy}}
\def\mjy{\hbox{\kern 0.20em mJy}}
\def\Mj{\hbox{\kern 0.20em MJy}}
\def\jy{\hbox{\kern 0.20em Jy}}
\def\ghz{\hbox{\kern 0.20em GHz}}
\def\srmd{\hbox{\kern 0.20em sr$^{-1}$}}

\def \mum{$\mu$m}
\def\G{\hbox{\kern 0.20em G}}

\def\htwo{\hbox{H${}_2$}}
\def\h13cop{\hbox{H$^{13}$CO$^{+}$}}

\def\h2o{\hbox{H$_2$O}}

\usepackage[T1]{fontenc}
\usepackage{ae,aecompl}

\usepackage{amsmath}
\usepackage{hyperref}

\graphicspath{{./}{figures/}}

\title{Modeling the early mass ejection in  jet-driven protostellar outflows. Lessons from Cep\,E. 
}
\titlerunning{Modeling the Cep\,E jet-driven outflow}
\authorrunning{Rivera-Ortiz et al.}

\begin{document}

   \author{
          P. R. Rivera-Ortiz\inst{1,2}
          \and
          A. de A. Schutzer \inst{2} 
          \and
           B. Lefloch\inst{2,3}
          \and
          A. Gusdorf \inst{4,5}
          }

   \institute{
   Instituto de Radioastronom\'ia y Astrof\'isica, Universidad Nacional Aut\'onoma de M\'exico, P.O. Box 3-72, 58090, Morelia, Michoac\'an, Mexico\\
    \email{p.rivera@irya.unam.mx}
   \and
   Univ. Grenoble Alpes, CNRS,
Institut de Planétologie et d’Astrophysique de Grenoble (IPAG), 38000 Grenoble, France
\and 
Laboratoire d’Astrophysique de Bordeaux, Univ. Bordeaux, CNRS, B18N, allée Geoffroy Saint-Hilaire, 33615 Pessac, France
    \and
    Laboratoire de Physique de l’ENS, ENS, Universit\'e PSL, CNRS,
Sorbonne Universit\'e, Universit\'e de Paris, 75005 Paris, France
    \and
    Observatoire de Paris, PSL University, Sorbonne Universit\'e,
LERMA, 75014 Paris, France
             }

   \date{Accepted January 12, 2022}

\abstract{
Protostellar jets and outflows are an important agent of star formation as they carry away a fraction of  momentum and energy, which is needed for gravitational collapse and protostellar mass accretion to occur.}{
Our goal is to provide constraints on the dynamics of the inner protostellar environment from the study of the outflow-jet propagation away from the launch region.}{
We have used the {axisymmetric} chemo-hydrodynamical code {\sc Walkimya-2D} to numerically model and reproduce the physical and CO emission properties of the jet-driven outflow from the intermediate-mass protostar CepE-mm, which was observed at $\sim 800$~au resolution in the CO $J$=2--1 line with the IRAM interferometer. Our simulations take into account the observational constraints available on the physical structure of the protostellar envelope. }{ 
{\sc Walkimya-2D} successfully reproduces the main qualitative and quantitative features of the Cep\,E outflow and the jet kinematics, naturally accounting for their time variability. Signatures of internal shocks are detected as knots along the jet. In the early times of the ejection process, the young emitted knots interact with the dense circumstellar envelope through high-velocity, dissociative shocks, which strongly decrease the CO gas abundance in the jet. As time proceeds, the knots propagate more smoothly through the envelope and dissociative shocks disappear after $\sim 10^3\yr$. The distribution of CO abundance along the jet shows that the latter bears memory of the early dissociative phase in the course of its propagation. Analysis of the velocity field shows that the jet material mainly consists of gas entrained from the circumstellar envelope and accelerated away from the protostar at 700~au scale. As a result, the overall jet mass-loss rate appears higher than the actual mass-ejection rate  by a factor $\sim 3$. }{Numerical modeling of the Cep\,E jet-driven outflow and comparison with the CO observations have allowed us to peer into the outflow formation mechanism with unprecedented detail and to retrieve the history of the mass-loss events that have shaped the outflow.
}

\keywords{ 
stars: formation --- ISM: jets and outflows --- astrochemistry}

\maketitle


\section{Introduction}
\label{sec:intro}


Protostellar  jets  and  outflows  are  an  ubiquitous phenomenon of the star formation process from the early Class 0 to the late Class I phase, when the parental envelope is dissipated. They are  an  important  agent  of  star  formation  feedback,  which  affects  the gas's physical and chemical properties from cloud scale down to the central parental cocoon, where mass accretion is occurring. On the one hand, they act as a source of energy and momentum into the parental cloud, and
they disperse the material of the parental core, directly impacting the star formation efficiency and the final stellar mass  (see \citet{Frank2014} for a review); on the other hand, they are thought to remove a significant fraction of the angular momentum from the star–disk system, enabling the gas in the accretion disk to reach the central protostar \citep{Konigl00}.

A natural consequence of the propagation of such high velocity outflows through the protostellar envelope and the ambient molecular medium are shock fronts \citep{Reipurth99}. 
These shocks both heat and compress the gas, which favors chemical reactions in the gas phase while they modify the dust grain properties and release a fraction of their material in the gas phase through, for example, sputtering and shattering, thereby leading to a different chemical composition than observed in the ambient, preshock gas. Commonly, the high relative abundance of CO and its low energy J transitions makes it a good tracer for outflows in the conditions of cold molecular clouds. {Even more}, the morphology, velocity, and sizes of a molecular outflow depend on the observed tracer, the luminosity, mass, and age of the outflow, which indicates its evolution and its inner structure \citep{Bally2016}. 

Over the years, there have been intense numerical efforts to simulate the chemical and dynamical structure of protostellar outflows and their evolution in the ambient interstellar gas. 
\citet{Smith1994}and \citet{Smith1997} carried out 3D numerical simulations of dense molecular jets drilling through a molecular environment. They succeeded in reproducing the morphology of the ``classical" CO bipolar outflows, which image the accumulated and accelerated cool gas. They showed that the simulations predict infrared shock structures remarkably similar to those found in highly collimated Class 0 outflows. They proved that in spite of the limitations in their description  of the physical conditions (dust, magnetic field), these hydrodynamical models could provide a meaningful description of the dynamics of  bipolar outflows from young stars, accurate enough images for comparison with observations, and the outflow mass and energy contribution to the interstellar medium. 

\citet{Raga1990}  showed that simulations with an arbitrarily imposed ejection velocity variability lead to the formation of chains of internal working surfaces traveling down the jet flow. Such features strikingly resemble Herbig-Haro (HH) flows, with a chain of aligned knots close to the outflow source, and a large “head,” resulting from the “turning on” of the jet flow at larger distances.
Later on, the authors showed that a two-mode ejection velocity variability leads to the formation of chains of “short period knots”, which catch up with each other to form “long period knots”. 
This class of models has successfully accounted for the properties of HH flows 
and  they nowadays dominate the literature on the theory of astrophysical jets 
\citep{Canto1985, Raga1990, Raga2003, Noriega-Crespo2014,  Frank2014, Rabenanahary2022}. On the other hand, most of the hydrodynamical codes developed so far have not included a chemical network and thereby do not address the molecular gas composition, in contradiction with the observational evidence that outflows from Class 0 protostars are often chemically active \citep[][]{Bachiller2001,Arce2007,Lefloch2017,Ospina-Zamudio2018,Desimone2020}.  

It is only with the advent of the first generation of chemo-hydrodynamical codes that it is now possible to accurately model the physical and chemical outflowing gas evolution, and to build spectroscopic diagnostics which can be tested against the  observational constraints provided by the large (sub)millimeter arrays such as the Northern Extended Millimetre Array (NOEMA) and the Atacama Large Millimeter/submillimeter Array (ALMA), and to investigate in detail the processes that determine the jet-outflow properties in order to quantify the energetic and chemical feedback of newly born stars on their environment and eventually on galaxies.
 
In this work, we use {\sc Walkimya-2D}, a new 2D hydrodynamical code coupled to a reduced gas phase chemical network allowing for the evolution of the CO chemistry in molecular outflows to be followed \citep{CRETAL18}. The advantage of this code is that it computes the time evolution of chemical species in a full gas dynamic simulation. The new functionalities of {\sc Walkimya-2D} allow us to investigate more thoroughly the dynamics and the physical structure of protostellar outflows. 

\begin{figure}[!th]
\begin{center}
\includegraphics[clip,trim={3.5cm 0 3.5cm 0},width=0.9\columnwidth]{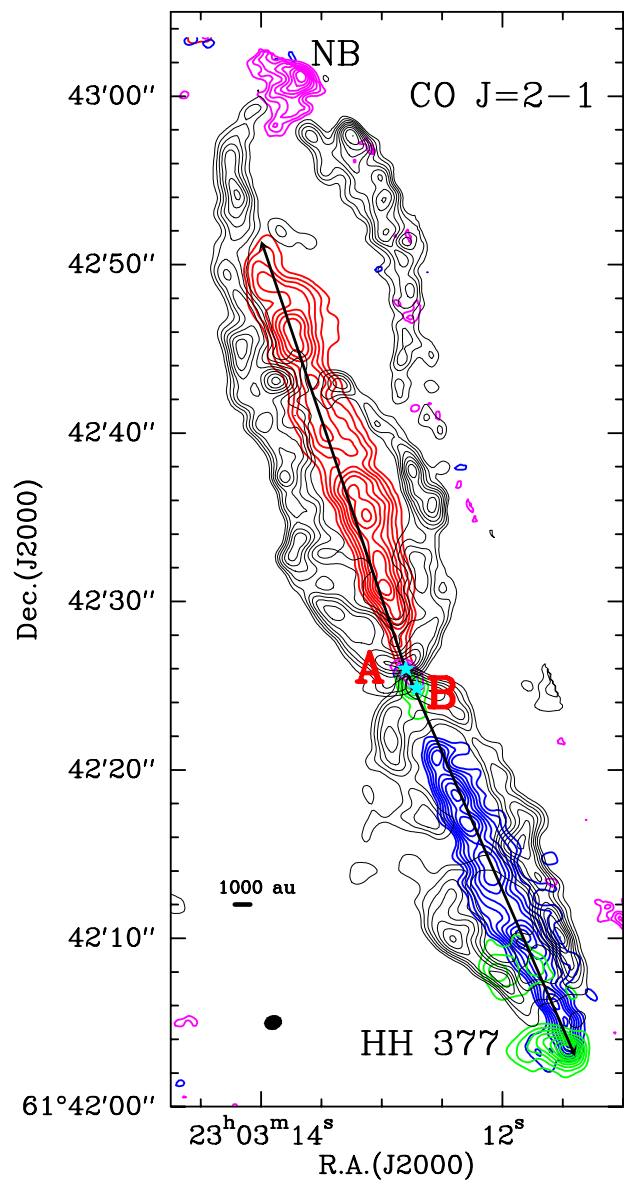}
\caption{CO $2$--$1$ line emission in the Cep\,E-mm outflow as observed with the PdBI at $1\arcsec$ resolution \citep{Lefloch2015}.
Four main velocity components are detected: a)~the outflow cavity walls (black) emitting at low velocities, in the range [$-8$;$-3$]$\kms$ and [$-19$;$-14$]$\kms$ in the northern and southern lobe, respectively; b)~the jet, emitting at high velocities in the  range [$-135$;$-110$]$\kms$ in the southern lobe (blue) and in the range [$+40$;$+80$]$\kms$ in the northern lobe (red); c)~the southern terminal bow shock HH377, integrated in the range [$-77$,$-64$]$\kms$ (green); d)~the northern terminal bullet NB integrated in the velocity range [$+84$,$+97$]$\kms$ (magenta). First contour and contour interval are $20\%$ and $10\%$ of the peak intensity in each component, respectively. The synthesized beam  ($1\farcs07 \times 0\farcs87$, HPBW) is shown in the bottom left corner. The main axis of the northern and southern lobes of the jet are shown with black arrows. \citep{Schutzer2022}}.
		\label{fig:CepE-CO}
	\end{center}
\end{figure}

 We have selected the high-velocity outflow associated with the intermediate-mass Class 0 protostellar system  Cep\,E-mm, located in the Cepheus OB3 association at a distance of $819\pm 16\pc$  \citep{Karnath2019}, whose CO emission observed at $1\arcsec$ with the Institut de Radioastronomie Millim\'etrique (IRAM) interferometer is displayed in Fig.~\ref{fig:CepE-CO}. This outflow has a luminosity of $\approx 100\lsol$ \citep{Lefloch1996} and a core mass of $35\msol$ \citep{Crimier2010}.  The source has been the subject of several detailed studies at arcsec angular resolution, in particular in the CO rotational transitions in the (sub)millimeter domain \citep{Lefloch1996,Lefloch2015,Ospina-Zamudio2019}, which have constrained the jet and outflow dynamical parameters (mass, mass-loss rate, momentum, density, temperature). The recent study by \citet{Schutzer2022} has brought a detailed view of the jet structure and its time variability, showing a complex interaction with the ambient protostellar material. It is also  one of the few protostellar jets for which a  full 3D picture of the gas kinematics is available. Overall, this outflow appears as an excellent testbed for chemo-hydrodynamical codes, with the prospect of getting more insight into the jet and outflow formation process.   
 
The goal of this study is to obtain an accurate numerical model of the Cep\,E molecular outflow in agreement with the constraints provided by the large (sub)millimeter observatories. In particular,  we aim to understand the dynamics of the jet and envelope interaction near the source.
The synergy between the {\sc Walkimya-2D} numerical simulations and the observations with the IRAM interferometer has allowed us to peer into the outflow formation mechanism with unprecedented detail
and to retrieve the history of the mass-loss events that have shaped the outflow.


This paper is organized as follows: in Section 2, we describe the numerical setup and the boundary conditions, which were chosen in agreement with the observational constraints on Cep\,E and present the best fitting model. The following sections present our main results: the outflow formation in Section 3, the gas acceleration mechanism in Section 4, the mass-loss history in Section 5, and the CO emission in Section 6. Finally, we summarize our conclusions in Section 7.


\section{Simulations}
\label{sec:2}

\subsection{{\sc Walkimya-2D}}

In order to study the temporal evolution of the molecular composition of a gas, one needs to solve the rate of change of the abundances of the different species contained in it. It is required to construct a chemical network gathering the creation and destruction reactions for the different species.
We have used the 2D chemo-hydrodynamical {\sc Walkimya-2D} that solves the hydrodynamic equations and a chemical network, on an axisymmetric numerical adaptive mesh. A complete description of the code is presented in \cite{CRETAL18}. 
Initially, both the jet and the surrounding quiescent protostellar gas have a  CO abundance of  1.6$\times 10^{-4}$ relative to H$_2$.
The chemical network is designed mainly to follow the evolution of the CO chemistry, as induced by the  heating-cooling processes in the jet propagation and the shock interaction(s) with the ambient gas. For computational reasons (to keep computational times within reach), the network is reduced to 14 chemical species, including C, O, H$_2$O, OH, and CO \citep{CRETAL18}. The reaction rates were obtained from the UMIST database \citep{McElroy2013}. 
 {\sc Walkimya-2D} and its chemical network were successfully benchmarked against laboratory experiments to study the NO formation by electrical discharges using both a zero-dimensional and a gas dynamic model approach \citet{CRETAL18}. Also, in the astrochemical context, the code was successfully benchmarked against the model of dark molecular cloud presented in \citet{McElroy2013}, and has been used to explain the CO emission produced by the Orion Fingers in the Orion KL region \citep{RGETAL22}.

The energy loss rate is calculated adopting the same prescription as in \citet{RO19a} (see references therein): for temperatures larger than 5800 K the cooling function considers the atomic contribution while in the lower temperature range, it uses a parametric molecular cooling function based on CO and $\htwo$.


The adaptative mesh uses seven levels of refinement, yielding $4096 \times 1024$ cells, in a computational domain ($5\times 1.25$)$\times 10^4$~au, allowing a resolution as high as 12.2~au per cell side. We used a reflective boundary condition for the symmetry axis and a free outflow boundary condition for all the other borders. The size of the mesh is large enough that the outer boundaries do not affect the simulation.  The jet is injected on the left side of the simulation box with the physical conditions indicated in Table~\ref{tab:models}.

\subsection{Initial conditions: the protostellar core}
 
In order to describe the protostellar core, we have adopted the physical conditions derived by \cite{Crimier2010}  from a simple 1D modeling of the dust spectral energy distribution between $24\mu m$ and $1300\mu m$, using the radiative transfer code 1D DUSTY \citep{Ivezic1997}. 
The temperature is fitted by  a broken radial power law $T\propto r^{\beta}$ with $\beta$= $-0.8$   in the range 50--$300\K$ and $\beta$= $-0.4$ in the range $7\K$--$50\K$.
In Crimier's model, the density profile is  assumed to start at a radius $r_{in}$= 70~au and to follow a single radial power law  distribution $n(r)$= $n_0 \times (r_0/r)^\alpha$.  

In order to avoid density singularities in the limit $r\to 0$ in the numerical modeling, we have adopted the slightly modified density profile
\begin{equation}
    n(r)=\frac{n_0}{1+ \left( r/r_0 \right)^\alpha },
\end{equation}
where $n_0=10^9$ cm$^{-3}$, $r_0=100$ au and $\alpha=1.9$. 

The simulation takes into account the role of gravity, in agreement with the source density stratification evidenced from Crimier's analysis. Our model includes the  resulting gravitational force as a source term, which reads: 
\begin{equation}
     g=-\frac{2\;c_0^2\;r}{r_0^2\left(1+\left(r/r_0\right)^\alpha\right)}    
\end{equation}
where $c_0$ is the local sound speed. The inclusion of this gravity term ensures { hydrostatic} equilibrium close to the source.


\subsection{Initial conditions: the jet}

The jet is bipolar and consists of a narrow (diameter < 400 au) central component surrounded by a collimated layer with a radial extent up to 1000 au, as shown by \citet{Schutzer2022}. The jet appears young, with a dynamical age of $1400\yr$, and compact, with a length of $\sim 0.1$--$0.14\pc$ (18000 -- 29000~au) for the southern and northern lobes, respectively. 
The physical conditions (temperature, \htwo\ gas density) in the outflow were estimated by \citet{Lefloch2015} and \citet{Ospina-Zamudio2019} from a CO multi-rotational line study complemented  with CO $J$=2--1 observations at $1\arcsec$ angular resolution with the IRAM interferometer. 
In the southern lobe, the jet appears to consist of a warm (T=$80$--$100\K$) gas component of  \htwo\ density $n(\htwo)$= (0.5--1.0)$\times 10^5\cmmt$ component and a higher-excitation component of  $n(H)$= (0.5--1.0)$\times 10^6\cmmt$ and temperature (T=$400$--$750\K$), which the authors associated with the high-velocity knots. Similar physical conditions are found in the northern lobe, with a kinetic  temperature T=$180$--$300\K$ and gas density $n(\htwo)$= (0.6--2.0)$\times 10^5\cmmt$. Therefore, the jet is rather massive, with a total mass of $0.03\msol$ ($0.09\msol$) in the southern (northern) lobe, after taking into account the revised distance to the source \citep{Karnath2019}.  

The jet asymptotic radial velocities were determined from the CO 2--1 line profiles as  $+65\kms$ and $-125\kms$ in the northern and southern lobe, respectively \citep{Lefloch2015}. The jet proper motions in both lobes were measured by 
\cite{Noriega-Crespo2014} from combining multiple mid-infrared IRAC $4.5\mu m$ observations and \htwo\ $2.12 \mu m$ images at a difference of 16 years. We note that the revised distance of Cep\,E-mm (820 pc  instead of 730 pc previously) does not significantly affect the tangential velocities, 
nor the inclination angle of the jet with respect to the plane of the sky derived by \cite{Lefloch2015}, which is now $47\deg$ ($40\deg$ previously estimated) and the estimated dynamical age of the jet (about $10^3\yr$). Based on these observations, the molecular jet velocity is estimated  $\sim 100\kms$  ($150\kms$) in the northern (southern) lobe, respectively.   
Kinematical analysis of the  molecular gas knots in the jet  shows evidence for time and velocity variability of the mass-ejection process in Cep\,E \citep{Schutzer2022}. 

In the simulation, the jet is launched at z=0, with a radius $r_j$, a gas density $n_j(H)$= $10^6\cmmt$, and a temperature $T_j$= $300\K$.
In order to account for knot formation inside the jet, we introduced variability in the gas injection, which we modeled following the equation
\begin{equation}
     V_j= V_{j,0} \left [1+\delta_v \cos(2\pi t /\tau) \right ]
\end{equation}
where  $\tau=130$~yr is the injection mass period and $t$ is the evolutionary time. The jet injection velocity $V_{j,0}$ and the relative amplitude variability $\delta_v$ are free parameters of the simulation. 
We adopted $\delta_v$= 0.05--0.08, which implies velocity variations of 10--$15\kms$, consistent with the typical variations reported by \citet{Schutzer2022} in Cep\,E and in other outflows such as IRAS04166+2706 \citep{Santiago-Garcia2009}.

\subsection{Comparison with observations}
\begin{table}[ht]
    \centering
    \begin{tabular}{l|c|c|c|c} \hline \hline
    Parameter    & M1 & M2 & M3 & M4 \\
    \hline
        $r_j$[au]                 & 100 & 50& 50 & 150 \\
        $v_{j,0}$ [km s$^{-1}$]   & 200 &165 &200  &200 \\
        $\delta_v$                & 0.08 & 0.08 &0.05 & 0.08\\
        \hline
    \end{tabular}
\caption{Initial parameters of the four models M1--M4 computed with {\sc Walkimya-2D}: jet radius $r_j$, injection velocity $v_{j,0}$, relative velocity variability amplitude $\delta_v$. {M1} is the model which best fits the Cep\,E outflow. }
    \label{tab:models}
\end{table}

We have run four models M1--M4, whose initial jet parameters are listed in Table \ref{tab:models}. For the sake of simplicity, our {\sc Walkimya-2D} simulations do not include the effect of jet precession.  Comparison between numerical simulations and observations are made with the northern outflow lobe of Cep\,E, whose entrained gas dynamics is better revealed in CO interferometric observations \citep{Schutzer2022}: morphology, gas acceleration, time-variability, and knot formation.

The simulations give us access to the physical structure of the outflow and its chemical properties. Filtering the emission of the high- and low-velocity components allows direct comparison between the simulations and the observational signatures of the distinct outflow components and therefore it is possible to quantify the physical processes involved in the outflow formation. 
In order to compare our numerical simulations with the CO observations, the emissivity has been computed directly from the hydrodynamical simulations assuming local thermodynamic equilibrium (LTE). We have constructed synthetic CO J= 2--1 maps of the outflow cavity and the jet, by integrating the emission in the velocity intervals [3;$7\kms$] and [50;$150\kms$], respectively. Those maps were subsequently convolved with a gaussian profile whose size (FWHM) corresponds to the synthetic beamsize  of the interferometer ($1\arcsec$ or 820~au). 

We first explored the parameter space and searched for the model which best reproduces the qualitative and quantitative properties of the Cep\,E northern outflow lobe. Overvall, we found that all four models M1 -- M4 show similar results. Therefore, in what follows we present the results of model M1, that is, the simulation that best accounts for the molecular gas observations of Cep\,E, whose 
initial jet parameters are summarized in Table~\ref{tab:models}. The initial jet velocity is $200\kms$, with a velocity fluctuation amplitude of 0.08, and an ejection radius of 100~au. 


\label{sec:results}
\begin{table}[]
    \centering
    \begin{tabular}{l|c|c|c|c|c|c} \hline\hline
Parameter    & M1 & M2 & M3& M4& \multicolumn{2}{c}{Cep\,E} \\
             &    &    &   &  & \multicolumn{2}{c}{North South} \\
    \hline
$M_{j}\, [10^{-2}$M$_\odot]$ &2.7   & 0.75& 1.5 & 5 & 3.6  & 2.6\\
$M_{o}[$M$_\odot]$           &2.3 & 1.3 & 1.2 & 2.9  & 2.0  & 0.4 \\
$z_{bs}[\,10^3$au$]$         &28  & 23 & 29  & 31   & 28   & 21\\
$r_{max}[\,10^3$au$]$        &4.8  & 1.5 &3.6 & 5.3  & 5  & 4 \\
        \hline 
    \end{tabular} 
    \caption{Outflow physical properties derived from CO emissivity maps at 1500~yr in the four numerical simulations (M1-- M4) and comparison with the observational values of the Cep\,E outflow: jet mass $M_j$, outflow mass $M_{o}$, jet length $z_{bs}$, maximum outflow radius $r_{max}$. These values are taken from \citet{Lefloch2015} and \citet{Ospina-Zamudio2019} and have been corrected for the revised distance to the source.}
    \label{tab:results}
\end{table}

\label{sec:bestfit}
\begin{figure*}[!ht]
    \centering
    \includegraphics[angle=0,width=0.85\textwidth, height=1.2\textwidth]{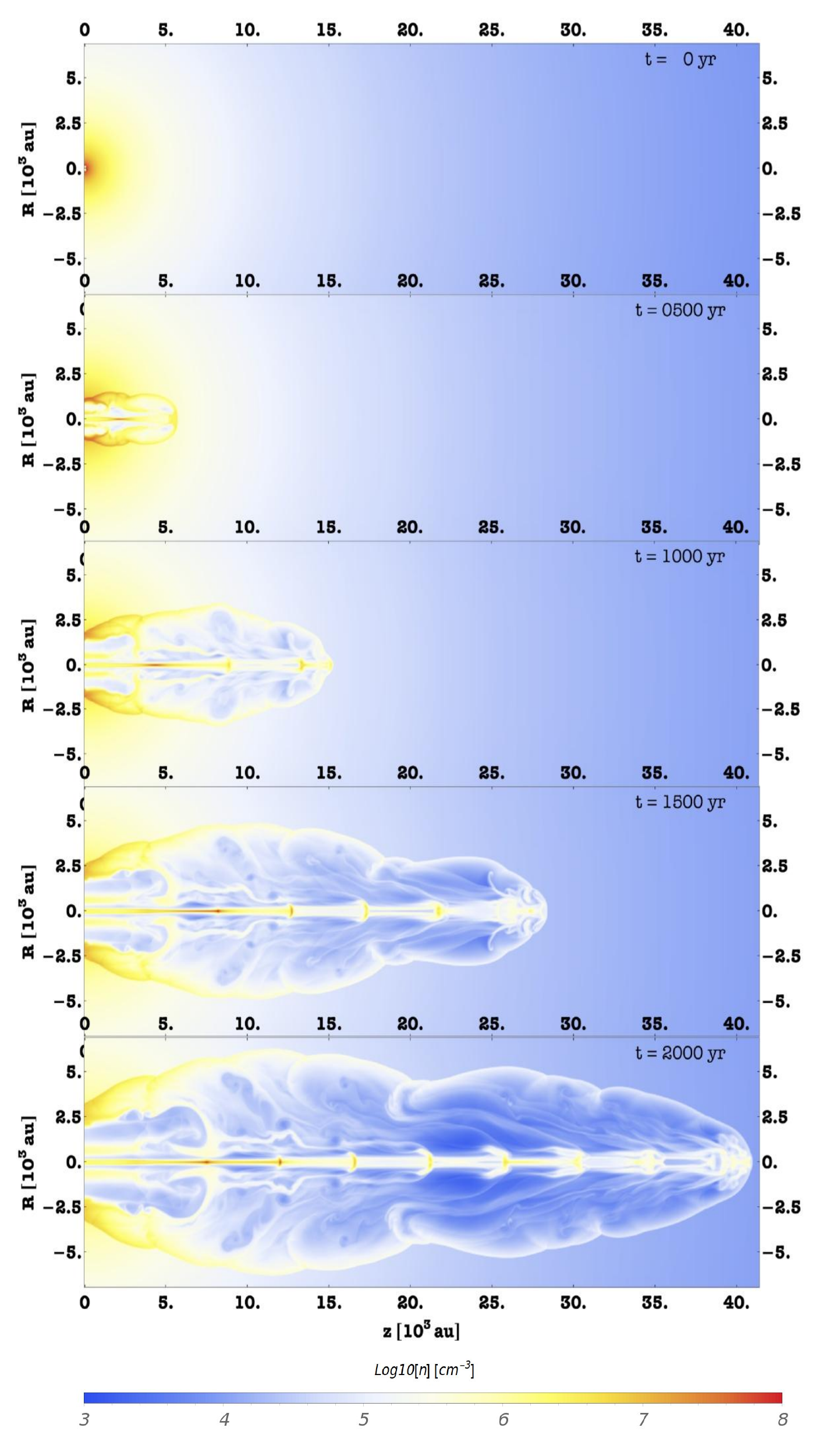}
    \caption{Outflow formation and impact of the jet on the gas density distribution as a function of time. Snapshots for 0, 500, 1000, 1500, and 2000 yr. The injection velocity variability forms internal working surfaces, or knots, and a structured cavity. The leading bowshock accumulates the mass of several knots.}
    \label{fig:jet2000}
\end{figure*}

We compare the numerical results to the main observational features of the northern outflow lobe, identified in the recent study by \citet{Schutzer2022}: morphology, gas acceleration, time-variability, and knot formation. 


\section{Outflow formation}
Figure~\ref{fig:jet2000} displays the molecular gas distribution at five timesteps of the outflow formation, from $t$=0 to $t$= $2000\yr$ by steps of $500\yr$. (A video is available at Fig.~\ref{fig:jet2000}). 

The jet is launched at $t$=0 into gas of density $10^9\cmmt$. The density contrast between the jet ($10^6\cmmt$) and the ambient gas ($10^9\cmmt$) makes the shock propagate into the ambient gas at a velocity much lower than the injection velocity $v_{j,0}$, an effect that has been studied in detail already by \citet{Raga1990}. 
Turning on the jet creates a high-velocity, collimated narrow component which propagates in the  ejection direction. It is surrounded by a cocoon of gas that expands laterally, creating a low-velocity (0--$10\kms$) “cavity" component, with dense walls surrounding the jet. The first working surface interacts with the very dense gas close to the jet source, which makes it decelerate, causing the second ejected knot to  catch up with the first bowshock. As shown in Fig.~\ref{fig:jet2000}, this process repeats a few times until the bowshock leaves the inner envelope and propagates into the protostellar gas, reaching its terminal velocity. It is only after $\sim 200\yr$ and the launch of 3 knots that the jet manages to drill out the inner 1000~au of the  protostellar envelope and to propagate into the surrounding gas (second panel from top).
As a consequence the overpressure resulting from the accumulation of knots pushes the protostellar envelope laterally away from the jet, causing the formation of a wide-angle outflow cavity, and a density increase in the low-velocity cavity walls {at the base of the jet}. This implies that a kinematic age computed from the bowshock position and velocity is actually a lower limit to the real age of the outflow.

As time proceeds, the model shows a leading jet head
and a series of traveling “internal working surfaces”
(IWS). These structures form as the result of the ejection time-variability and are observed as "knots", or overdensity regions with a small size of a few $\sim 100$~au. At $t$= $1000\yr$, 3 knots are easily detected along the jet, while at $t$= $2000\yr$, 8 eight knots are easily identified. It is worth noticing they tend to expand radially as they propagate along the jet, tracing the wings of inner bowshocks, since the slower material interacts with the faster material ejected at later times. These wings appear to drive the formation of complex structures inside the outflow cavity as can be seen in the bottom panel of Fig.~\ref{fig:jet2000}. Fast moving knots tend to accumulate at the head of the bow, increasing the density at the tip. After $1500\yr$, the knots located at distance $> 15000$~au display wings with a typical size of 1000~au. 

One expects the gas density to decrease as a result of its radial expansion in the course of the jet  propagation. However, looking in detail at the jet density field at e. g. $t$= $1000\yr$ (second panel in Fig.~\ref{fig:jet2000}), it appears that the gas density does not decrease monotonically along the narrow jet. The density distribution of fast material along the propagation direction is analyzed in Section 5.2. 

At $t$= $1500\yr$, a second outflow cavity has now formed. The first outflow cavity has reached a size of $\approx$ 15000~au and expanded up to a radius of about 5000~au ($6\arcsec$ at the distance of Cep\,E-mm), and the head of the outflow has reached a distance of $2.8\times 10^4$~au ($34\arcsec$). At later times,  the second cavity expands to reach a similar radius of $\approx 5000$~au. 

Inspection of Figure~2 reveals that a complex network of relatively dense structures ($n \sim 10^5$--$10^6\cmmt$) forms inside the outflow cavity very early. One can note the presence of "filaments" along the jet in the first $1000\yr$.
As can be seen in the panel at $t$= $1000\yr$, a filament has formed along the jet. Located at $R= 500$ au, it is detected up to $z= 2000$ au and appears as a thin, yellow, wiggling structure parallel to the jet.   
We also note the presence of a "shell", or a bow, which formed in the dense protostellar envelope and moves slowly ($\sim 20\kms$) as it propagates over 8000au in $2000\yr$, whereas the jet reached $40\times 10^3$~au. This dense shell (or bow) connects the jet envelope of entrained gas to the cavity walls.

Our numerical results on the outflow morphology appear in good agreement with the NOEMA observations of the Cep\,E northern lobe (Fig.~1). As can be seen in Table~\ref{tab:results}, both the length  and the radius of the outflow cavity are correctly reproduced; this good match between observations and simulations is obtained for a computational timescale of $1500\yr$, similar to the estimated Cep\,E dynamical age ($1400\yr$).  Even more, the jet momentum  obtained from the simulation is $3.78 $ M$_\odot$ km s$^{-1}$, comparable to the momentum obtained by \cite{Lefloch2015} of $2.5 $ M$_\odot$ km s$^{-1}$, taking into account the inclination angle. 

In order to better understand the role of the core initial conditions in shaping the outflow morphology, we also carried out simulations without the inclusion of the high-density core given by Crimier nor the related gravitational term. In these cases, it turned out that the low-velocity outflows formed in the ambient medium of the jet propagation were actually far too collimated with respect to what is observed, in particular close to the protostar and the jet launch region. Therefore, it appears that the wide opening of the low-velocity outflow cavity depends on the density of the dense inner region and the contribution of the latter to the gravitational field, two parameters that are related to the evolutionary stage of the system. The importance of including a density distribution and a gravitational term is in agreement with \citep{RaCa93} and \citep{Cabrit1997} who proposed that a steep radial density decrease would produce a wider opening angle for jet-driven outflows.

\section{Gas acceleration}

\begin{figure}[!ht]
\centering
\includegraphics[clip,trim={0cm 8cm 0cm 8cm},angle=0,width=\columnwidth]{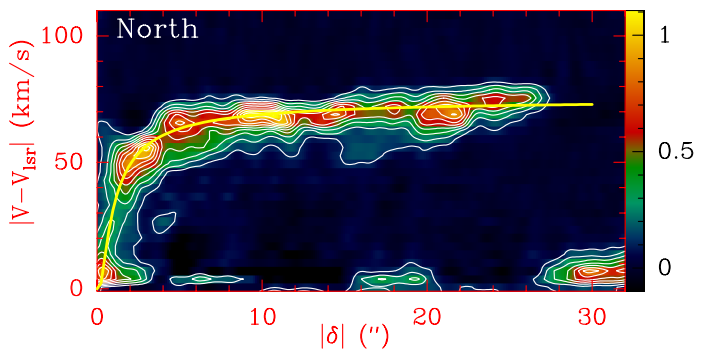}
\caption{Cep\,E northern outflow lobe. Position-Velocity diagram of the CO J= 2--1 line emission along the jet main axis. Positions are in arcsec offset relative to the location of the driving protostar Cep\,E-A. First contour and contour interval are 10\% of the peak intensity. We have superimposed in yellow the best fitting solution $(V-V_{lsr})$/$V_0$= $\exp(-\delta_0/\delta)$ where $\delta_0$= 690~au \citep{Schutzer2022}.}
    \label{fig:pv-main}
\end{figure}

\begin{figure}[!ht]
     \centering
    \includegraphics[clip,trim={2cm 0 3cm 0},width=0.28\textwidth,angle=-90]{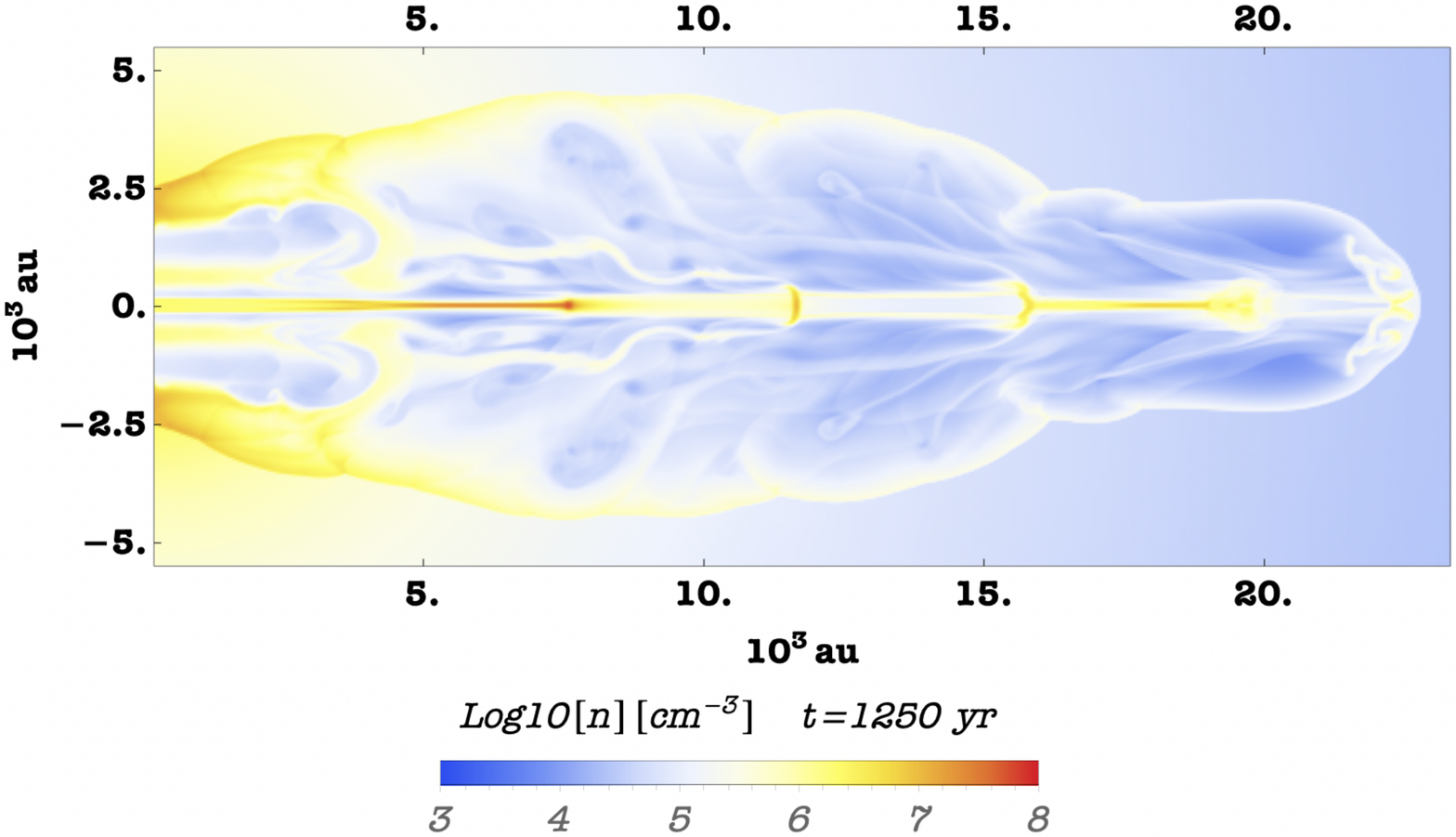}
    \centering
    \includegraphics[width=0.48\textwidth]{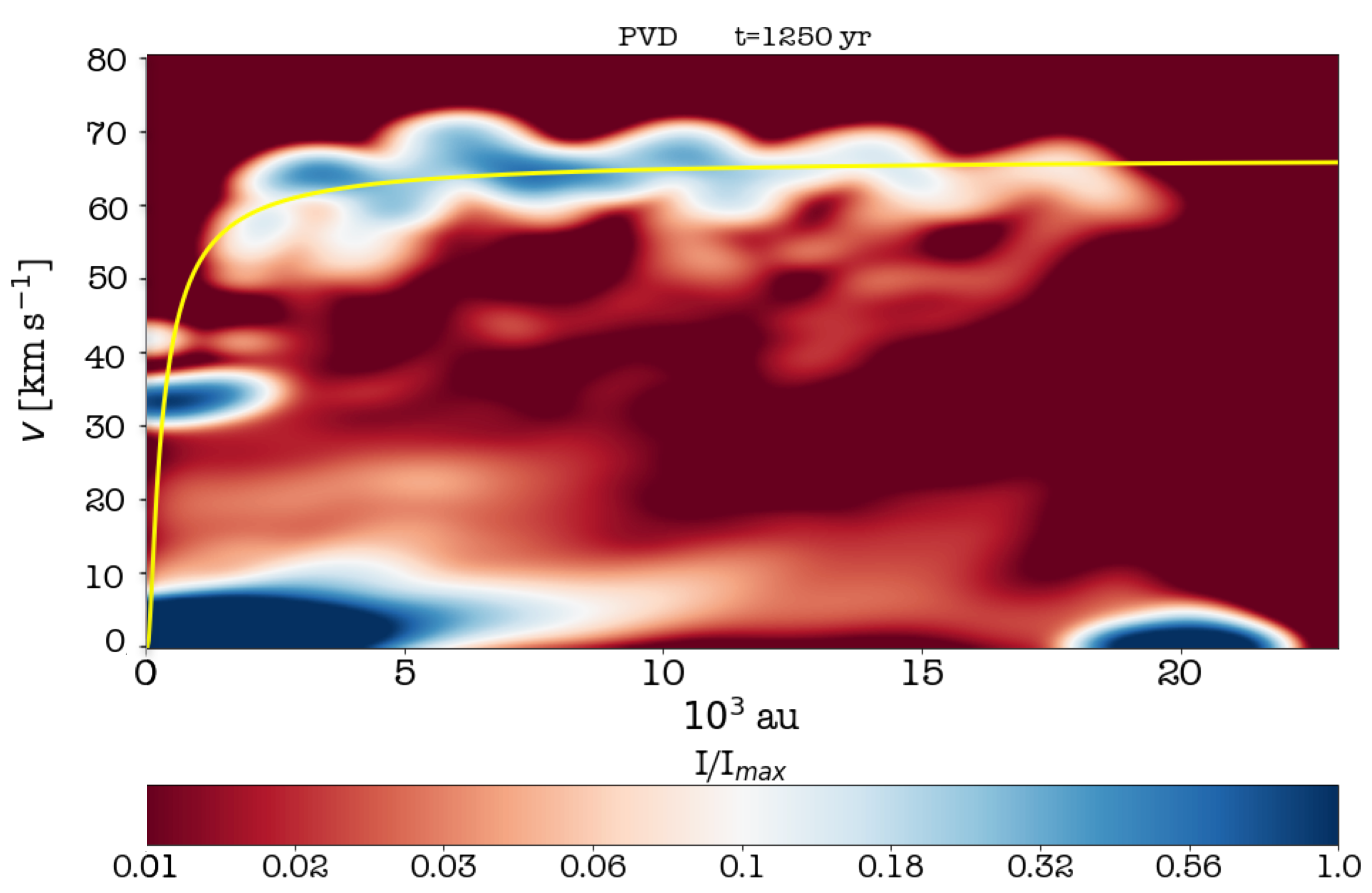}
    \caption{~Best fitting model {(M1)} to the Cep\,E northern outflow. (top)~Density distribution at $1250\yr$. The jet has propagated over a distance of $2.4\times 10^4$~au and formed a low-velocity cavity of 3000~au radius. Five knots and their associated wings are detected along the jet.(bottom)~Position-Velocity diagram of the CO emission along the jet main axis obtained {using the $47^\circ$ outflow inclination angle with respect to the plane of the sky. }
    } 
    \label{fig:M4-n-pv}
\end{figure}

\begin{figure}[ht]
\centering
\includegraphics[clip,trim={0cm 8cm 0cm 8cm},angle=0,width=\columnwidth]{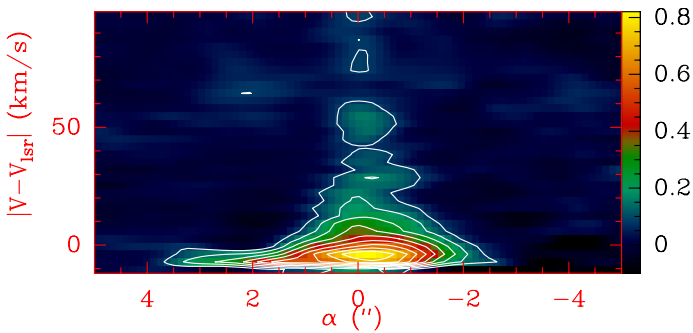}
\caption{Cep\,E northern outflow lobe. Position-Velocity diagram of the CO J= 2--1 line emission across the jet main axis at $\delta$= $0\arcsec$. Positions are in arcsec offset relative to the location of the driving protostar CepE-A. First contour and contour interval are 10\% of the peak intensity \citep{Schutzer2022}.}
   \label{fig:pv-minor}
\end{figure}

Cep\,E  displays evidence of gas acceleration along the jet over a distance of 1000~au away from the protostar (Fig.~\ref{fig:pv-main}). This effect was reported and observationally characterized by \citet{Schutzer2022}.
We show in the bottom panel of Fig.~\ref{fig:M4-n-pv}, the synthetic Position-Velocity diagram of the CO gas along the main axis of the jet. Two kinematical components are clearly identified: i)~a low velocity component, between $+0$ and $+10\kms$ up to 8000~au ($10\arcsec$) from the source, associated with the outflow cavity walls; ii)~a high velocity component, that displays a periodical behavior, reflecting the mass injection variability and the internal shocks, where the fast material catches up with the slower material. The jet is detected at a radial velocity close to $V_j$= $65\kms$ at $z > 2000$~au from the source.

Our simulations show that the effective velocity of the outflow is significantly lower than  the injection jet velocity. This appears as a result of the  jet interaction with the dense central envelope. In this case, the Position-Velocity diagram in Fig.~\ref{fig:M4-n-pv} shows that the CO gas emission accelerates from ambient velocity 
($V$= $0\kms$) to the terminal jet velocity on a short scale $< 5\times 10^3$~au. Close to the source, gas structures accelerated up to $V\sim 50\kms$ are also detected along the jet axis up to 1000--2000~au (bottom panel of Fig.~\ref{fig:M4-n-pv}). 

An observational signature of envelope gas can also be found in the Position-Velocity diagram of the CO $J$=2--1 emission across the jet main axis, as can be seen in Fig.~\ref{fig:pv-minor}. 
The CO emission contours show how the ambient material initially at rest at $V_{lsr}$= $-11\kms$ is gradually accelerated as one gets closer to the location of the protostar at $\alpha$= $0.0\arcsec$. In addition to the jet mainly detected up to  $V\sim 60\kms$, signatures of high-velocity knots (up $+90\kms$)  are  also identified close to the protostar (Fig.~\ref{fig:pv-minor}). 

{It is worth remembering that the synthetic Position-Velocity diagrams use a wide slit in the direction of the projected jet axis and the contribution of all the possible velocity projections along the line of sight, using an angle of $47^\circ$ with respect to the plane of the sky,} whereas the Cep\,E observational plots (Fig.~\ref{fig:pv-main}) display the radial component of the jet velocity.
For the sake of comparison with the Cep\,E jet, we have checked whether the solution found by \citet{Schutzer2022} ($V-V_{lsr})/V_{j}$= $\exp(-\delta_0/\delta)$ could provide a reasonable fit to the synthetic jet velocity profile along the main axis. The yellow curve drawn in the bottom panel of  Fig.~\ref{fig:M4-n-pv} traces the best fitting solution of Schutzer et al. obtained for a length scale $\delta_0$= 690~au, applied to the radial jet velocity $V_r$=$ 65\kms$. Taking into account the inclination of the jet with respect to the line of sight, the agreement between numerical and observational jet velocity profiles is very satisfying both qualitatively and quantitatively. 

To conclude, our simulation satisfyingly accounts for the acceleration of material from the protostellar envelope by the Cep\,E jet, from ambient velocity up to reaching the radial terminal jet velocity $V_{j}\sim 90\kms$. This process occurs over a scale of 5000~au.

\section{Mass-loss history}
\subsection{The jet}

\begin{figure}[ht]
\centering
\includegraphics[angle=0,width=\columnwidth]{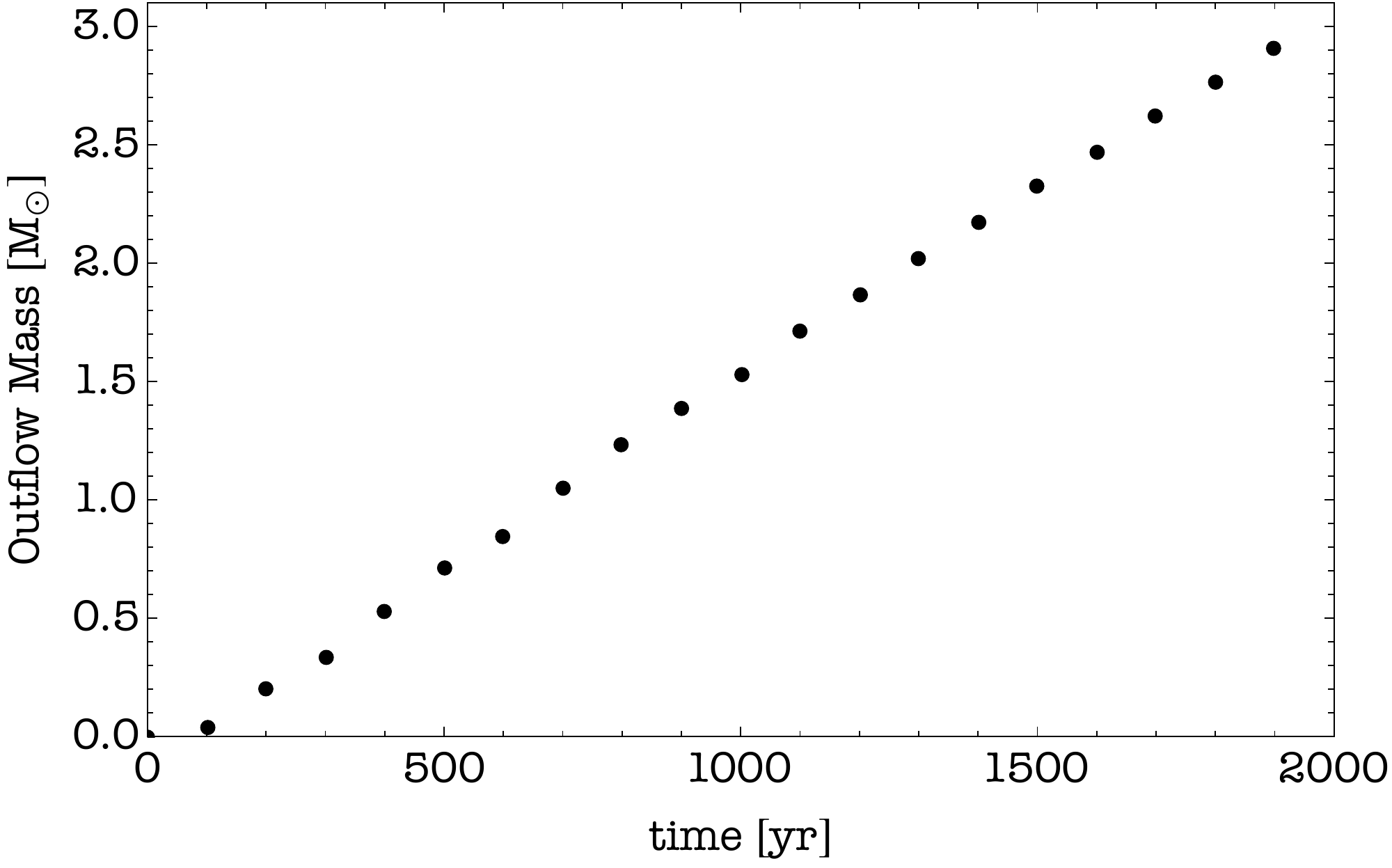}
\caption{Model {M1}. Outflow mass evolution as a function of time.}
\label{fig:outflowmass}
\end{figure}

We first measured the mass of outflowing material as a function of time. The mass integration was performed over the whole velocity range. The result is displayed in Fig.~\ref{fig:outflowmass} and shows that the outflow mass increases linearly with time at a rate of $1.6\times 10^{-3}\msol\yrmu$. At a time of $1500\yr$ in the simulation (similar to the dynamical age of Cep\,E) the mass of outflowing gas amounts to $2.3\msol$, a value in good agreement with the observational determination in the northern lobe (see Table~\ref{tab:results}).

\begin{figure}[ht]
\centering
\includegraphics[angle=0,width=\columnwidth]{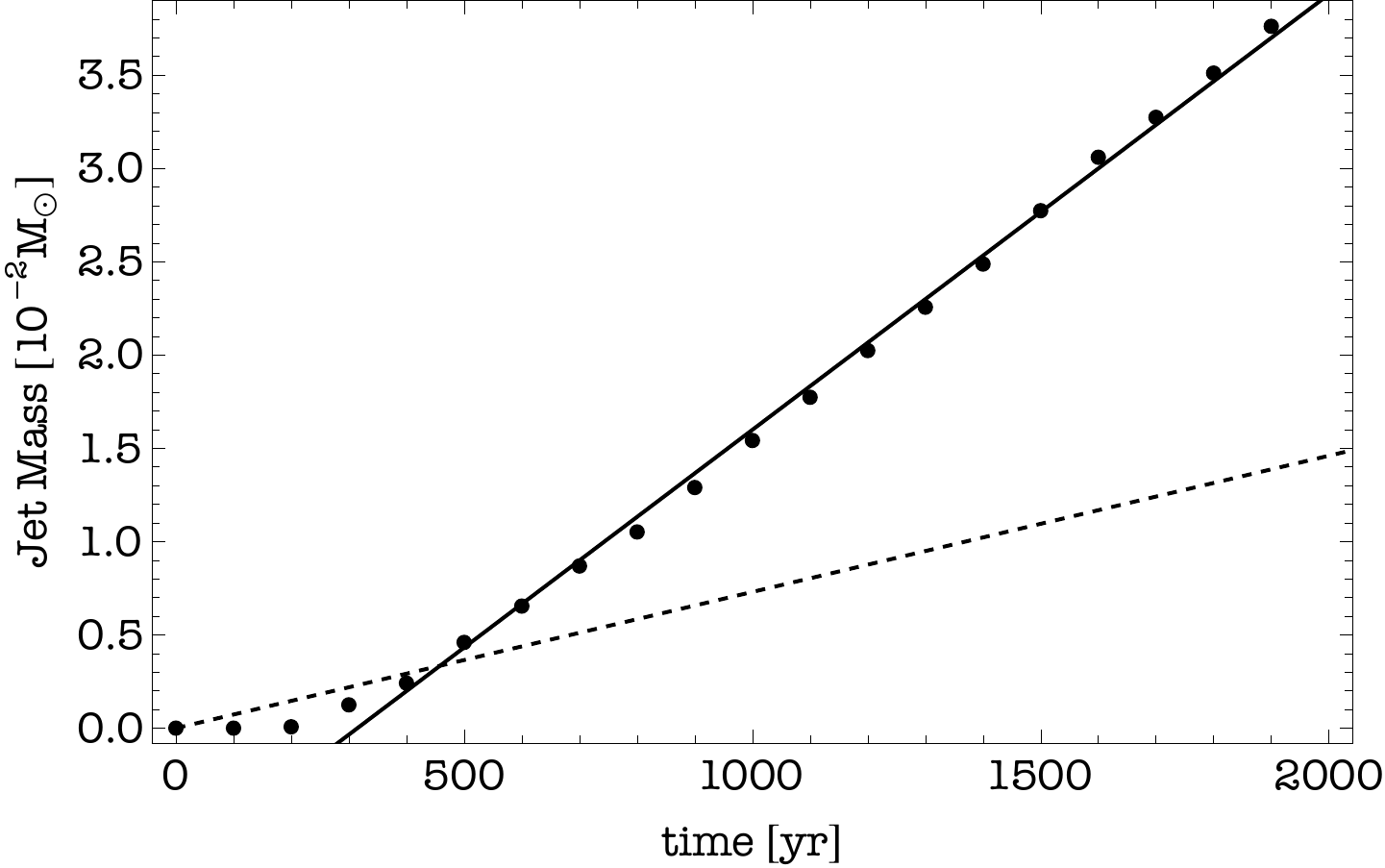}
\caption{Model {M1}. Jet mass evolution as a function of time. The points are obtained from the simulation every 100 yr. The black continous line is a linear fit of the points starting from 300 yr, with a slope of 2.3 $\times 10^{-5}M_\odot$ yr$^{-1}$ and the dashed line corresponds to the material injection rate of  $\dot{M}=7.3\times10^{-6}$ used in the simulation.}
\label{fig:jetmass}
\end{figure}

In a second step, we have studied the variations of the jet mass as a function of time. We
took into account all the gas at velocity $V > 50\kms$ to estimate the jet mass and follow its variations. The variations of the jet mass as a function of time are displayed in Fig.~\ref{fig:jetmass}. 
After an initial delay of $\sim 200\yr$, which corresponds to the time needed for the knots to drill the envelope, the jet mass increases linearly with time, at a rate $\dot{M}$= $2.3\times 10^{-5}\msol\yrmu$ in very good agreement with the observational determination by \citet{Schutzer2022} ($2.7\times 10^{-5}\msol\yrmu$). As discussed previously in Sect.~3.2, the protostellar material appears to feed the high-velocity jet through an entrainment process. 
An observational signature of this effect is found in the Position-Velocity diagram of the CO $J$=2--1 emission across the jet main axis, displayed in Fig.~\ref{fig:pv-minor}. 
The CO emission contours show how the ambient material initially at rest at $V_{lsr}$= $-11\kms$ is gradually accelerated as one gets closer to the location of the protostar at $\alpha$= $0.0\arcsec$. In addition to the jet mainly detected up to  $V\sim 60\kms$, signatures of higher-velocity knots (up $+90\kms$)  are  also identified close to the protostar (Fig.~\ref{fig:pv-minor}). 

Third, in order to quantify the magnitude of the entrainment effect to the jet mass, we have disentangled the contributions of the injected material from the entrained material.
Theoretically, the jet mass injection rate per unit of time can be estimated as a function of the injected material density $n_{j}$, the jet velocity $v_{j}$ and the jet injection radius $r_{j}$: \\
\begin{equation}
   \left[ \frac{\dot{M}}{{\rm M}_\odot / \rm{yr}}  \right] =
    1.5\times {10^{-6}}  
    \left[ \frac{r_{j}}{50 \rm{au}} \right]^2
    \left[\frac{n_{j}}{10^6\cmmt} \right]  
    \left[\frac{v_{j}}{165\kms}\right].
    \label{eq:mdot}
\end{equation}

Numerically,  the jet mass injection rate in the computational domain is  in reasonable agreement with the simple, theoretical description from Eq.~\ref{eq:mdot} ($\dot{M}=7.3\times 10^{-6} \msol\yrmu$ with the conditions for model M1). In both numerical and theoretical cases,  the jet mass injection rate is lower than the  jet mass rate $\dot{M}$ of the simulation by a factor of $\approx 3$. We consider the difference between both values as very significant as it highlights the importance of the entrainment process of circumstellar material in the formation of molecular jets.

\subsection{The knots}
We now examine the physical properties of the knots which form along the jet in the numerical simulation and their evolution with time. We then compare them with their observational counterparts in the Cep\,E jet.

\subsubsection{Formation and evolution}

\begin{figure}[!ht]
    \centering
    \includegraphics[clip,trim={6cm 0 6cm 0},width=0.9\columnwidth,height=0.85\textheight]{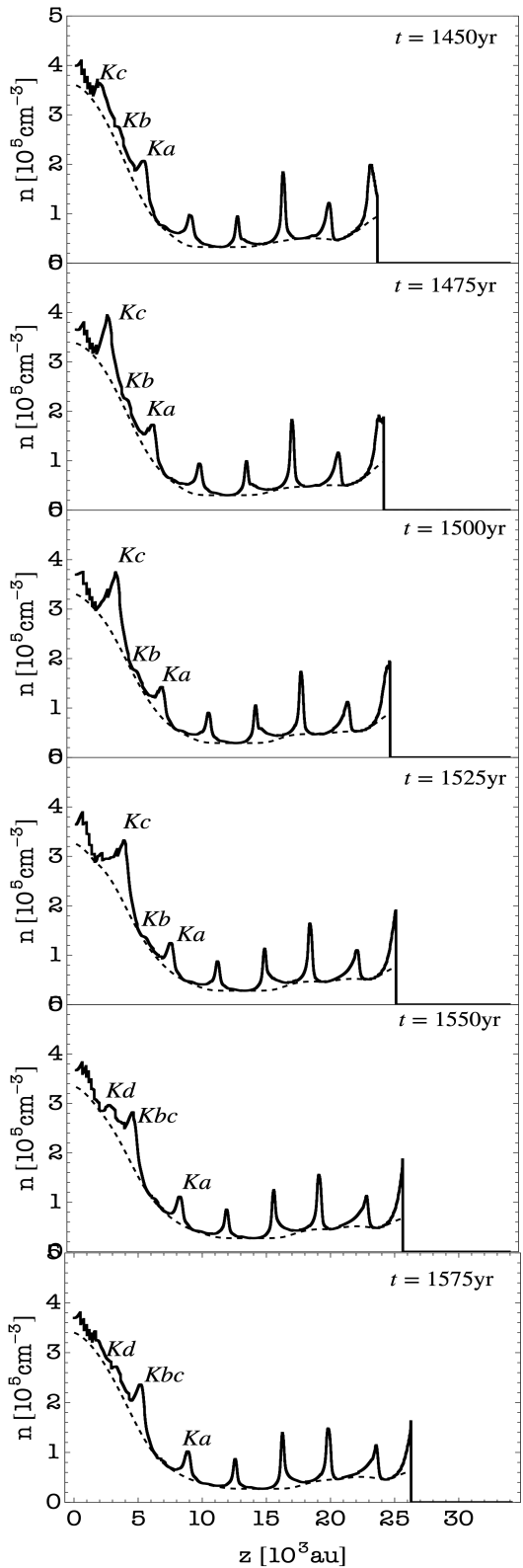}
    \caption{Gas density distribution along the jet main axis
    for different snapshots from 1450 to 1575 yr (dashed lines). Knots can be identified as overdensities in the density distribution (thick lines) in the order of $10^5$ cm$^{-3}$. To follow their behavior some knots have been labeled as Ka, Kb, Kc and Kd.
    }
    \label{fig:knotsposition}
\end{figure}

We have studied the gas density distribution along the jet as a function of time in the simulation up to $2000\yr$. We found that the evolution is characterized by very short timescales on the order of $25\yr$. In order to illustrate and better explain the process at work, we present here in  Fig.~\ref{fig:knotsposition} a montage of the jet density profiles at intervals of $25\yr$ over a time of $125\yr$, between $1450\yr$ and $1575\yr$. 
The mean density distribution appears steady and constant at distances further than $10^4$ au, while at shorter distances,  there is a decreasing density profile. This is a fossil signature of the steep initial density profile of the envelope, since it is expected to be difficult to remove the dense envelope material close to the source, even when it has been processed by the countinuous injection of the jet.
The peaks, which appear on top of the mean density distribution, result from the ejection velocity variability in the jet source (Sect.~2.3), which causes fast material to shock with lower velocity gas. Each of these peaks corresponds to a local overdensity structure in the jet, whose properties were derived from a gaussian profile fitting, leading to  a typical size of 1000~au, a density of about $10^5\cmmt$, and a typical mass of $10^{-3}\msol$. The physical properties of these  peaks are actually very  similar to those of the CO knots reported by \citet{Schutzer2022} in the Cep\,E jet, and, for that reason, they are probably their counterparts in the numerical simulation. 
Based on Fig.~\ref{fig:knotsposition}, it appears that at distances larger than $10^4$~au the density  and the knot distributions are almost steady, which reflects the fact that the knots are  propagating into the low-density material of the jet at a speed of about 525~au in $25\yr$. The high density in the inner $10^4$~au makes the situation drastically different. \citet{RO19a} have shown that gas flowing into an environment of similar density can create reverse shocks that propagates at a significant fraction of the original velocity relatively to the shock velocity. This is precisely the present situation with a jet of density $10^6\cmmt$  propagating into gas of density 2--$4\times 10^5\cmmt$. This effect is seen in Fig.~\ref{fig:knotsposition}, where the number of peaks varies in the inner $10^4$au. As an example we follow the formation of {the knot labeled Kbc}: at $t=1450\yr$ Kc looks as a short and wide peak, that gets taller and thiner up to $1500 \yr$, when it appears to have merged with Kc. Immediatly after this, a small peak Kd forms behind it at $1550\yr$, which moves forward at a slightly lower velocity, creating a very wide peak and giving a similar density distribution at 1450 yr and 1575 yr, an interval of $125\yr$, which is approximately the period of variability used in the simulation. This process gets repeated several times,  accelerating the fossil envelope material, as described in Sec. 4, relating the process of gas acceleration with the formation of knots.

\begin{figure}[!ht]
\centering
\includegraphics[width=\columnwidth]{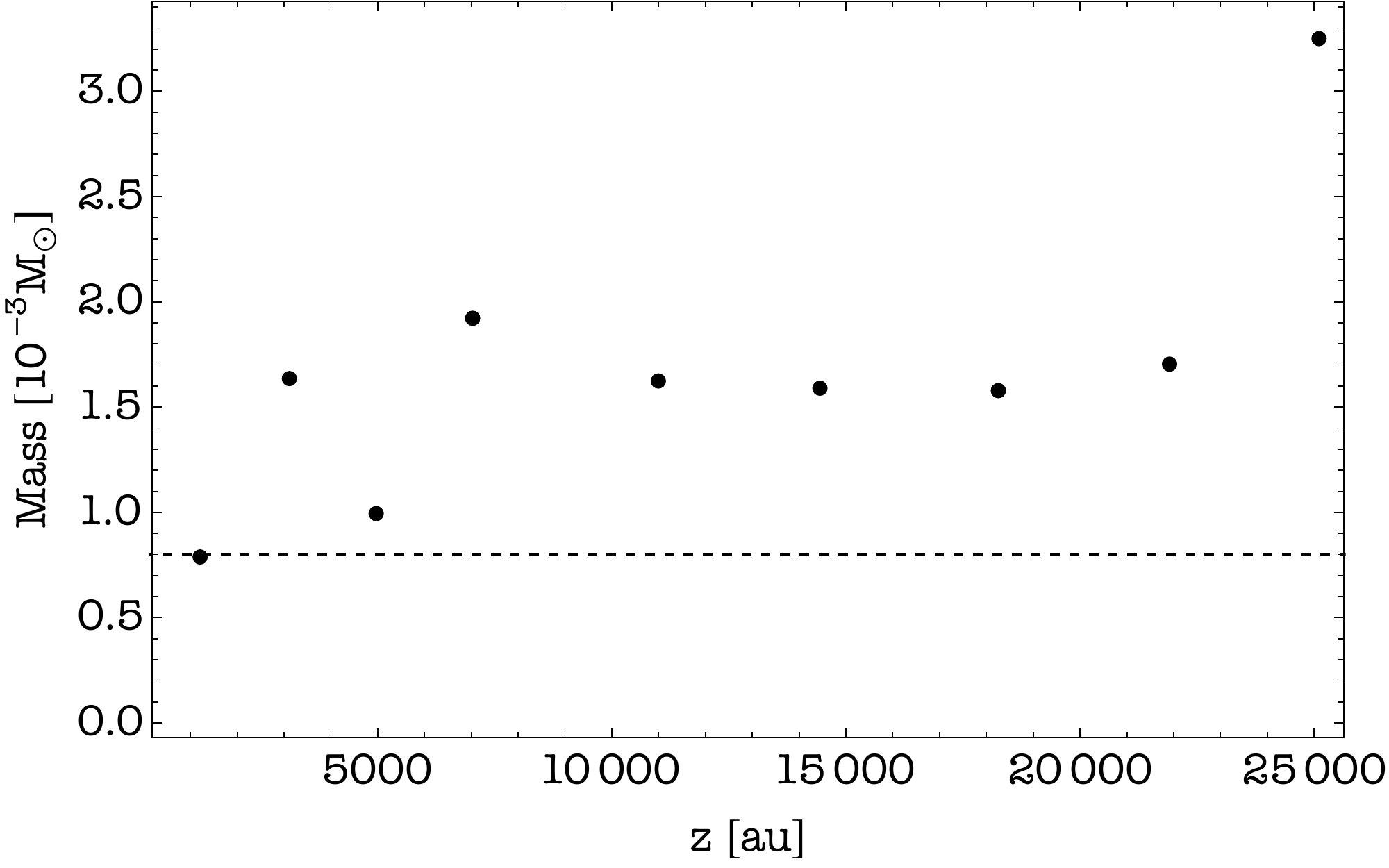}
\caption{Mass of the knots detected at 1500 yr in Model M1 as a function of their position $z$ along the jet, projected by 47$^\circ$, which corresponds to the angle between the Cep\,E jet main axis and the plane of the sky. The dashed line corresponds to the mass injected to the simulation in a single period.}
    \label{fig:massknots}
\end{figure}

We have reported in Fig.~\ref{fig:massknots} the mass of the knots as a function of their location along the jet at $t$= $1500\yr$. It appears that the knots tend to increase their mass by a factor of 2 as they move away from the injection site. The simulation shows that the mass increase occurs on a very short length scale as the second knot at $z\sim 3000$~au has already a mass of $1.7\times 10^{-3}\msol$. Beyond 8000~au, all knots display rather similar and steady masses, $\sim 1.7\times 10^{-3}\msol$. The terminal knot is an exception due to its higher mass ($> 3\times 10^{-3}\msol$). This high value is consistent with the fact that several knots have already reached the tip of the jet and are accumulating there, as can be seen in panels at t=1500 and $2000\yr$ in Fig.~\ref{fig:jet2000}.

\subsubsection{Observational comparison}

In the later times of the simulation, it appears that the spatial separation between the knots located close to the source is actually shorter by about a factor 0.5 than the separation between those located in the older region of the jet. This is consistent with the bimodal distribution of dynamical ages reported by \citet{Schutzer2022}.


The physical parameters of the knots (size, mass, density) such as derived from the numerical simulation, are therefore in good agreement with the observations. \citet{Schutzer2022} noticed that the distribution of knot dynamical ages appeared to be bimodal, with the presence of two timescales:  a short time interval of 50--$80\yr$ close to the protostar (8000 au), corresponding to the first 4--5 knots, and a longer timescale (150–-$200\yr$) at larger distances from the protostar. 

\begin{figure}[!ht]
    \centering
    \includegraphics[width=\columnwidth]{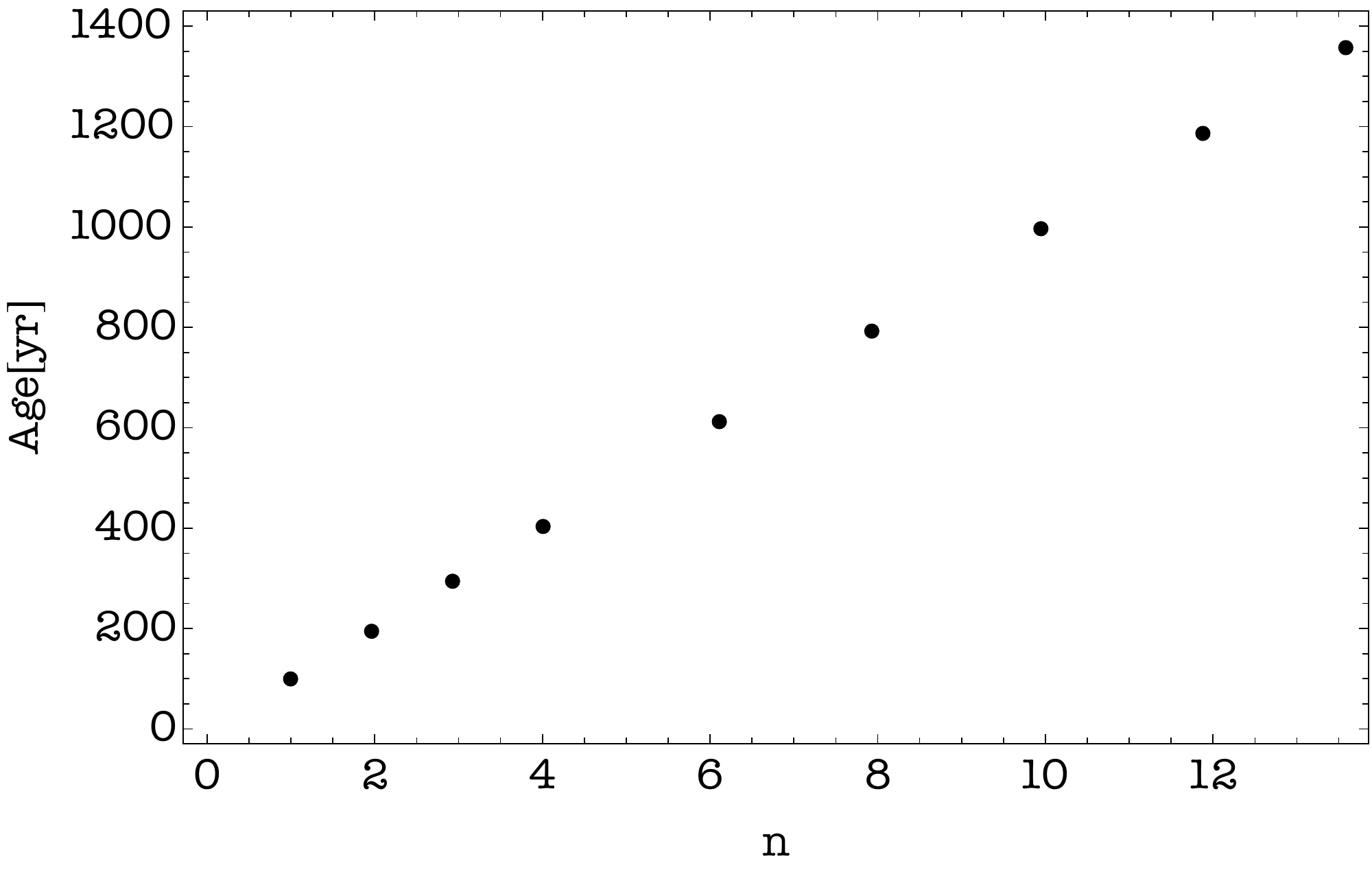}
    \caption{Dynamical age of the knots identified in model {M1} at $t$= $1500\yr$ in terms of a multiple number $n_i$ of the shorter dynamical age. }
    \label{fig:ageknots}
\end{figure}

Interestingly, the numerical simulation reports the same trend, as can be seen in Fig.~\ref{fig:ageknots}, in which the ages of all the knots identified along the jet have been reported. The first four younger knots are separated by a timescale of about $100\yr$, whereas the following knots are separated by a timescale of about $200\yr$, or twice the value measured close to the jet launch region. Our simulations show that a high-velocity knot can overcome the spatial separation to the previous ejecta, which was slown down as a result of the interaction with the dense gas of the protostellar envelope. The typical timescale for the collision is $500\yr$ (Fig.~\ref{fig:ageknots}).

We propose that the bimodal
distribution is actually the signature of the interaction of
the knots with the high-density gas inside the shell. As soon
as the knots run out of the shell and propagate into lower
density gas, they encounter free motion conditions.

\section{CO emission}

\begin{figure}[!ht]
\centering
\includegraphics[width=0.49\textwidth]{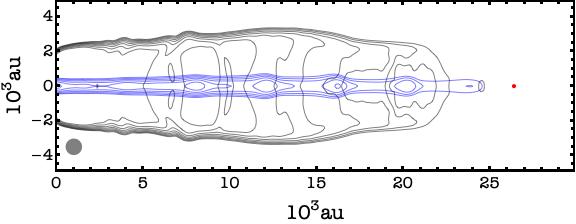}
\caption{{Synthetic map of Model M1 at $t$= $1500\yr$ {considering a projection angle with respect to the plane of the sky of 47$^\circ$, as observed in Cep\,E-mm }}. CO J= 2--1 integrated emissivity contour maps of the cavity (Black contours, 3 to 7 km s$^{-1}$) and the jet (Blue contours, 50 to 150 km s$^{-1}$) components. Scaling is logarithmic. The response of the IRAM interferometer was modeled by a gaussian of 830 au diameter (FWHM), corresponding to a beam size of $1\arcsec$ (FWHP) at the distance of Cep\,E-mm, which is drawn with a gray disk (bottom left). The location of the frontal shock is  marked with a red point.}
    \label{fig:COem}
\end{figure}

Thanks to the detailed modeling of the CO chemistry by {\sc Walkimya-2D}, we could produce synthetic CO integrated emissivity maps of the outflow jet and cavity, assuming LTE and optically thin emission. As an example, Fig.~\ref{fig:COem} displays the CO emissivity of the low-velocity outflow cavity integrated between 3 and $7\kms$ (black contours) and of the high-velocity jet, integrated between 50 and $150\kms$ (blue contours), as calculated by Model M1  at $1500\yr$. 
The CO image reveals the previously identified highly collimated structure of the jet with a typical width of $\sim 1000$~au, which slightly increases with distance from the source. 
An important result is that the CO emission both from the high-velocity (50-$150\kms$) jet (blue contours) and the low-velocity (3-$7\kms$) cavity (black contours) drops $\sim 2000$~au before the terminal bowshock, whose location is marked by a red spot in the plot.  The lack of low-velocity CO emission over a few 1000 au near the apex of the outflow cavity is consistent with the terminal bowshock being very efficient at dissociating the ambient molecular gas.

\begin{figure}
    \centering
    \includegraphics[width=\columnwidth]{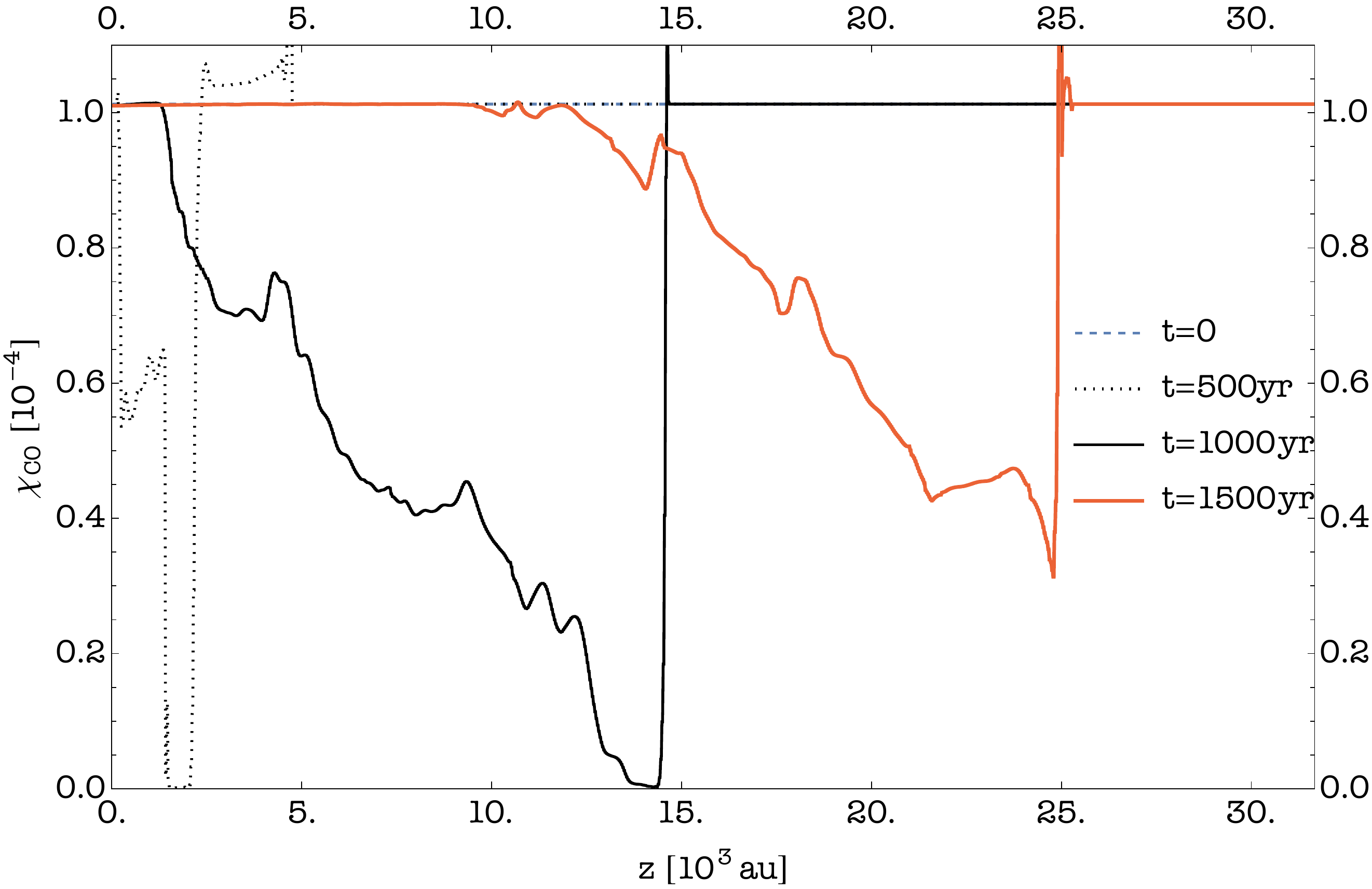}
    \caption{{\bf Model M1}. Variations of the CO/$\htwo$ abundance relative to its initial value ($10^{-4}$) along the high-velocity jet at 4 different times of the simulation: 0, 500 (dotted), 1000 (solid black), $1500\yr$ (solid red). }
    \label{fig:XCO}
\end{figure}
In order to gain more insight into the time variations of the CO abundance (relative to $\htwo$)  in the outflow, we have computed the CO abundance distribution along the high-velocity jet at 4 different times of the simulation in model M1: 0, $500\yr$, $1000\yr$, and $1500\yr$. The  distributions are shown in Fig.~\ref{fig:XCO}. For the sake of simplicity, we have reported the ratio $\rm X_{CO}$ of the CO abundance to its initial value, taken equal to $10^{-4}$, a standard value in dense interstellar clouds \citep{Lacy1994}. 

In the very early stages of the simulation (t < $500\yr$), when the jet has run only a few $1000$~au, X$_{\rm CO}$ is well below its canonical value of 1.0. For instance, we measure $\rm X_{CO}\approx 0.5$ at $5000$~au from the protostar at $t=500\yr$. At later times in the simulation ($1000\yr$ and $1500\yr$ in model M1, see Fig.~\ref{fig:XCO}), two regimes are identified in the $\rm X_{CO}$ distribution observed along the jet: 
 The first regime is close to the jet launch region  $\rm X_{CO}$= 1. The length of this region increases with time (2000~au at t=$1000\yr$, 10000~au at t=$1500\yr$) as the jet propagates at  velocity V$_j$= $140\kms$.
The second regime, at a larger distance, $\rm X_{CO}$ decreases from 1 to a few $10^{-1}$ with increasing distance to the protostar. The decrease is rather smooth and does not display abrupt variations along the jet. The minimum value is reached at the terminal bowshock before $\rm X_{CO}$ jumps abruptly again to 1, which stands as the canonical value in the ambient, quiescent gas in which the jet propagates.

To summarize, our simulation shows that  CO is partly destroyed  in the early times of the jet launch, close to the protostar ($\rm X_{CO}$ < $10^{-1}$). The destruction process appears to stop $\approx 900\yr$ after the beginning of the simulation and the CO abundance of the ejected material remains steady, equal to its initial value. We propose that CO dissociation occurs as a consequence of the violent shocks caused by the first  high-velocity knots (and the jet) when they are launched and  impact the dense, protostellar envelope at the beginning of the simulation.  It is only when the knots finally drill out the envelope and escape out of the inner protostellar region, entraining a fraction of the ambient protostellar envelope, that the local value of $\rm X_{CO}$ tends to return to its initial value. This process can also be influenced 
by the gas entrainment process. 
 
Interestingly, we note that the slope of the $\rm X_{CO}$ distribution along the jet at $t$= $1000\yr$ and $t$= $1500\yr$ is similar, suggesting that it does not vary with time. Also, the velocity variations between subsequent knots ($\sim 10\kms$) do not appear to significantly affect the $\rm X_{CO}$ distribution. Therefore it seems that the jet kept the memory of the initial launch process in the $\rm X_{CO}$ axial distribution. 

This numerical result has several implications on the observational side. First, lower values of $\rm X_{CO}$ may actually reflect the conditions of the jet formation process. Of course, precession and knot interaction with the ambient gas (or the cavity) may alter this conclusion. We speculate that in the latter case, the $\rm  X_{CO}$ variations occur on a much smaller length scale, corresponding to the size of the knots, i.e. 1000~au typically.
Our model suggests that the  distribution of $ \rm X_{CO}$ along the jet could probably provide constraints robust enough to discriminate between the processes at work. 

On the observational side,  \citet{Gusdorf2017} studied the emission of the OI $63\mu m$ line at $\sim 6\arcsec$ resolution with the Stratospheric Observatory For Infrared Astronomy (SOFIA) and detected both the signature of the Cep\,E southern jet and the terminal bowshock HH377. A column density ratio N(OI)/N(CO)$\sim 2.7$ was measured in the jet, indicating that that the jet is essentially atomic in the region of HH377. 
The authors proposed that the OI emission could arise 
from dissociative J-type shocks or with a radiative precursor, 
caused by the knots propagating at a different velocity in the jet \citep{Lehmann2020}. Interestingly, no signature of OI was detected in the jet toward shock position BI, located halfway between the protostar and HH377. 

As discussed above, our numerical simulations propose an alternative explanation for the origin of OI in the Cep\,E jet. A detailed map of the OI emission along the southern jet with  SOFIA, would help to confirm whether the CO dissociation is localized only toward HH377 or if it is also  present along the jet, hence could trace the history of the early stages of the mass-ejection process. 








\section{Conclusions}
\label{sec:conclusions}

Using the reactive hydrodynamical code Walkimya-2D, which includes a chemical network based on CO\citep{CRETAL18}, we have carried out a set of simulations in order to reproduce the morphology and the physical properties of the molecular jet-driven outflow of the intermediate-mass protostellar source Cep\,E, as derived from previous observational studies  \citep{Lefloch2015,Gusdorf2017,Ospina-Zamudio2019,Schutzer2022}. Outflow precession was not considered in this work. 
We have obtained a very satisfying agreement, both qualitatively and quantitatively, with the observations when modeling a time-variable jet with initial density $n(H)$= $10^6\cmmt$, radius $r_j$= 100~au, temperature $T_j$= $300\K$, velocity $V_j$= $200\kms$ propagating into the density stratified protostellar envelope as modeled by \citet{Crimier2010}. Our main results are as follows: 
\begin{itemize}
\item~The jet and cavity morphologies (width and length) are consistent with the observations and the expected kinematics using a variable jet ejection $\delta V/V_j$= $0.08$. The best fitting solution is obtained at a numerical timescale 
$t_{dyn}\simeq 1500\yr$, consistent with the jet dynamical timescale estimated observationally \citep[$\sim 1400\yr$]{Schutzer2022}. 

\item~The jet terminal velocity ($\simeq 90\kms$) is different from the injection velocity $V_j$ ($200\kms$), as a result of the first ejections being decelerated by the dense envelope and the entrainment of a layer of ambient protostellar material. It implies that the jet dynamical timescale is actually lower than the duration of the ejection phase.  

\item~The jet acceleration reported observationally by \citet{Schutzer2022} is consistently reproduced in the simulations. It appears to be the result of ambient material entrainment by the knots on a length scale of $\sim 700$~au, in agreement with the observational data. 

\item~We reproduce the properties of the knots along the jet assuming a periodic ejection. 
We found evidence for knot interactions in the dense inner protostellar region, where the densities of the local gas and the jet are comparable, which lead to the formation of secondary shocks in the close protostellar environment. At larger distances from the protostar, the lower ambient gas density allows the knots to propagate freely, without significant interaction. We propose that this process could account for the bimodal distribution of knots observed along the Cep\,E jet.  
Knots have a typical size of 1000-2000~au, with a mass of $\sim 1.5\times 10^{-3}\msol$, and density of $10^5\cmmt$. The mass carried away by the knots in the jet translates into a steady ejection mass rate of $2.3\times 10^{-5}\msol\yrmu$, a factor of 3 higher than the mass injection rate into the jet ($7.3\times 10^{-6}\msol\yrmu$). This difference is the signature of the entrainment and the subsequent acceleration of ambient material by the jet.

\item~The shock interaction of the jet knots with the protostellar envelope in the early times of the simulation lead to the dissociation of CO.  The destruction process appears to stop after $\approx 900\yr$, time from which the CO abundance of the ejected material remains steady and equal to its initial (canonical) value. As a consequence, the older part of the jet is characterized by a  lower CO gas abundance, which gradually decreases as one moves closer to the jet head. 
\end{itemize}

Our simulations underline the importance of mass-ejection time variability in the molecular outflow formation process and its interaction with the protostellar envelope. More work should be done, both observationally and numerically, in order to investigate the role of knots and their importance in the dynamical evolution of other young protostellar systems. 
Interferometric observations at subarcsec angular resolution of Cep\,E and other young protostellar jets should be undertaken as they would allow to make a significant step forward by bringing extremely useful constraints on the knot internal structure  and the dynamical processes at work in the jet. 
In parallel, the high spatial resolution accessible in the {\sc Walkimya-2D} numerical simulations 
($\sim 10$au) gives access to a novel view on the structure of protostellar jets  and their dynamics, which will help interpret a new harvest of observations.
%

\section*{Acknowledgments}
PR-RO, AS, BL acknowledge support from a)~the European Union’s Horizon 2020 research and innovation program under the Marie Skłodowska-Curie grant agreement No 811312 for the project “Astro-Chemical Origins” (ACO); b)~the European Research Council (ERC) under the European Union’s Horizon 2020 research and innovation program for the Project “The Dawn of Organic Chemistry” (DOC) grant agreement No 741002; c)~the UNAMPAPIIT grant IN110722.  Some of the computations presented in this paper were performed using the GRICAD infrastructure (https://gricad.univ-grenoble-alpes.fr), which is partly supported by the Equip@Meso project (reference ANR-10-EQPX-29-01) of the programme Investissements d'Avenir supervised by the Agence Nationale de la Recherche and the Miztli-UNAM supercomputer, project LANCADUNAM-DGTIC-123 2022-1.





%

\vspace{5mm}

\end{document}